\documentclass[12pt,a4paper,final]{iopart}

\usepackage{iopams}  
\usepackage{graphicx}
\usepackage[breaklinks=true,colorlinks=true,linkcolor=blue,urlcolor=blue,citecolor=blue]{hyperref}

\newcommand{\boF}{\mathbf{F}}
\newcommand{\PI}{\mathbf{\Pi}}
\newcommand{\boPhi}{\mathbf{\Phi}}
\newcommand{\fo}{\mathbf{f}_0}
\newcommand{\go}{\mathbf{g}_0}
\newcommand{\ho}{\mathbf{h}_0}
\newcommand{\bE}{\mathbf{E}}
\newcommand{\bcalE}{\boldsymbol{\mathcal{E}}}
\newcommand{\bcalS}{\boldsymbol{\mathcal{S}}}

\newcommand{\bH}{\mathbf{H}}
\newcommand{\bcalH}{\boldsymbol{\mathcal{H}}}

\newcommand{\boGamma}{\mathbf{\Gamma}}

\newcommand{\bor}{\mathbf{r}}

\newcommand{\bp}{\mathbf{p}}
\newcommand{\bm}{\mathbf{m}}

\newcommand{\bcalP}{\boldsymbol{\mathcal{P}}}
\newcommand{\bcalM}{\boldsymbol{\mathcal{M}}}

\def\eps{\varepsilon}
\def\Re{\mathrm{Re}}
\def\Im{\mathrm{Im}}

\begin{document}

\title{Mechanical separation of chiral dipoles by chiral light.}

\author{Antoine Canaguier-Durand, James A. Hutchison, Cyriaque Genet, and Thomas W. Ebbesen}
\address{ISIS \& icFRC, University of Strasbourg and CNRS, 8 all\'{e}e Gaspard Monge, 67000 Strasbourg, France.}
\ead{genet@unistra.fr}

\begin{abstract}
 We calculate optical forces and torques exerted on a chiral dipole by chiral light fields and reveal genuinely chiral forces in combining the chiral contents of both light field and dipolar matter. Here, the optical chirality is characterized in a general way through the definition of optical chirality density and chirality flow. We show in particular that both terms have mechanical effects associated respectively with reactive and dissipative components of the chiral forces. Remarkably, these chiral force components are directly related to standard observables: optical rotation for the reactive component and circular dichroism for the dissipative one. As a consequence, the resulting forces and torques are dependent on the enantiomeric form of the chiral dipole. This suggests promising strategies for using chiral light forces to mechanically separate
chiral objects according to their enantiomeric form. 
\end{abstract}

\vspace{2pc}

\section*{Introduction}
Seminal historical works have emphasized how the notion of chirality is important in relation to light-induced motional effects, analyzing the mechanical interaction of particles with the angular momentum of a light beam \cite{PoytingPROC1909,BethPR1936}. Rotational movements due to the transfer of spin angular momentum from circularly (and linearly) polarized light to birefringent (i.e. chiral) crystals were observed long ago \cite{BethPR1936} and more recently at the level of birefringent micro-objects or particles and designed chiral micro-structures \cite{HigurashiAPL1994,HigurashiJApplPhys1997,FrieseNature1998,HigurashiPRE1999,EriksenAppOpt2003,liu2010light}. For absorbing particles, spin and orbital angular momenta (OAM) are transferred to the particles with the same efficiency, with rotation of the particles in circular orbits in the case of OAM \cite{HePRL1995,FriesePRA1996,ONeilPRL2002,SimpsonOptLett1997,DholakiaPRA2002,DariaJOpt2011,LehmuskeroNanoLett2013}. In all such experiments, both forces and torques are at play. First, for the majority of the setups, the particles are optically trapped with high numerical aperture objectives where gradient optical forces confine the particles in regions of highest intensities. Then, intrinsic optical torques are at play in relation with rotation induced by transfer of spin angular momentum, both at the level of birefringent and absorbing particles. Finally, radiation pressure is involved in OAM transfers that move a single particle around a circular orbit by the action of a tangential force driving the particle around a optical vortex \cite{DholakiaPRA2002}. \\
\indent In parallel, a new area in optics has emerged through the discovery of exotic light modes with complex field topologies~\cite{Grier2003revolution,Dennis2009singular,Dogariu2013optically}. Such light fields provide a unique framework to demonstrate new light-matter interactions, at the classical and quantum levels~\cite{Allenbook,Jain2012quadrupole,Lembessis2013enhanced}. Intriguing opto-mechanical effects have been proposed lately in relation to optical negative forces and tractor beams, fueling much effort and debate \cite{Chen2011tractor,Saenz2011tractor,albaladejo2009scattering,bliokh2013dual,bekshaev2013subwavelength,Canaguier2013force,Grier2013comment}. In this context, extensive discussions on the measurement of optical chirality have brought new concepts such as optical chirality density and flow to the forefront of optics, from fundamental and applied perspectives~\cite{LipkinExistence1964,TangOptical2010,BliokhCharacterizing2011,Cameron2012optical,Andrews2012chirality}. For instance, so-called `superchiral' fields provide fascinating perspectives in the context of ultra-high resolution spectroscopy of chiral molecules \cite{HendryUltrasensitive2010,TangEnhanced2011,ChoiLimitations2012}.  \\
\indent In this article, we propose a systematic description of the forces and torques that chiral light fields can exert on chiral objects. We probe fundamental couplings between chirality of light and chirality of matter in the dipolar regime. This regime affords us the clearest view of the physical mechanisms at play and of the underlying light-matter symmetries. We generalize to induced chiral optical forces the well-known decomposition of achiral forces into reactive and dissipative components \cite{cohen1992atom}. These two components are associated respectively, from the field point of view, with the chirality density and flow and, from the matter point of view, with optical rotation and circular dichroism. Because they involve optical chirality in its most general sense, our results go beyond recent studies that focused on light scattering and forces on chiral objects for linear or circular polarization states \cite{GuzatovQElec2011,ShangOptX2013,TkachenkoPRL2013,DingarXiv}. We reveal in particular new motional effects directly related to the enantiomeric form of the chiral objects, reminding that a chiral object can exist in two different forms, called enantiomers, that are non-superposable mirror images of one another. From a practical point of view, these effects constitute an original mechanism for enantiomeric separation based on chiral light forces. The feasibility of applying our mechanism in actual experiments is discussed precisely. 

\section{Chiral light fields}
We first consider arbitrary electromagnetic ($\bcalE,\bcalH$) fields in vacuum with an energy flow (i.e. Poyting vector) $\bcalS(\bor,t) = \bcalE \times \bcalH$ and an energy density $W(\bor,t) = \frac{\eps_0}{2} \left\| \bcalE \right\|^2 + \frac{\mu_0}{2} \left\| \bcalH \right\|^2$ -split into electric $W^{(E)}$ and magnetic $W^{(H)}$ parts- connected through a continuity equation $\partial_t W+\nabla\cdot\bcalS=0$ for the conservation of energy. It is also possible to characterize an associated optical chirality through a chirality density 
\begin{eqnarray}\label{chi_density}
K(\bor,t) =\frac{\eps_0 \mu_0}{2} ( \bcalH \cdot \dot{\bcalE} - \bcalE \cdot \dot{\bcalH} )  
\end{eqnarray}
and a chirality flow 
\begin{eqnarray} \label{chi_flow}
\boPhi (\bor,t) =\frac{\eps_0}{2}\bcalE \times \dot{\bcalE} + \frac{\mu_0}{2} \bcalH \times \dot{\bcalH} = \frac{\omega}{2} \left( \eps_0 \boPhi^{(E)} + \mu_0 \boPhi^{(H)} \right)
\end{eqnarray} 
where $\boPhi^{(E)}=\bcalE \times \dot{\bcalE}/\omega$ and $\boPhi^{(H)}=\bcalH \times \dot{\bcalH}/\omega$ are the electric and magnetic ellipticities. Both quantities are related too by a continuity equation $\partial_t K+\nabla\cdot\boPhi =0$ (see \cite{BliokhCharacterizing2011}).  These parameters characterizing the chirality of a light field will be derived for several simple examples in Section~\ref{section:simple_examples}. 

Light fields exert instantaneous force $\boF$ and torque $\boGamma$ on both electric and magnetic dipoles with \cite{bekshaev2013subwavelength}:
\begin{eqnarray}
\boF &= \left( \bcalP \cdot \nabla \right) \bcalE +\left( \bcalM \cdot \nabla \right) \bcalH + \mu_0 \dot{\bcalP} \times \bcalH - \eps_0 \dot{\bcalM}\times \bcalE  \nonumber \\
\boGamma &= \bcalP \times \bcalE + \bcalM \times \bcalH ~ . \label{f_t}
\end{eqnarray}

\section{Chiral dipoles}
A small chiral object, such as a small birefringent particle or a chiral molecule, interacting with ($\bcalE,\bcalH$) can be described as a coupled system of induced electric $\bcalP$ and magnetic $\bcalM$ dipole moments \cite{BarronBook}. In the linear and harmonic regimes, the system writes as
\begin{eqnarray}\label{chi_dip}
\bp_0 = \alpha \cdot\bE_0 + \imath \chi \cdot\bH_0 
\hspace{2cm} \bm_0 = -\imath \chi \cdot\bE_0 + \beta \cdot\bH_0 ~ ,
\end{eqnarray}
where $\bcalP=\Re[\bp_0(\bor) e^{-\imath \omega t}]$, $\bcalM=\Re[\bm_0(\bor) e^{-\imath \omega t}]$ and $\bcalE=\Re[\bE_0(\bor) e^{-\imath \omega t}]$, $\bcalH=\Re[\bH_0(\bor) e^{-\imath \omega t}]$. For simplicity here, we assume isotropic responses so that the complex polarizability tensors ($\alpha$, $\beta$, $\chi$) become complex scalars. \\
\indent Specific to chirality is the complex mixed electric-magnetic dipole polarizability $\chi$ \cite{TretyakovBook}. Its sign determines the $(+,-)$  enantiomeric form associated with the dipolar system (\ref{chi_dip}). Note that the electric and magnetic polarizabilities are invariant through enantiomeric changes since they are respectively quadratic forms of the electric and magnetic dipolar moments \cite{Craig1999new}. One aim of this article is to identify the effect of such a $\pm \chi$ sign change on optical forces exerted by light on a chiral dipole. The core of the discussion therefore consists in evaluating the force and the torque (\ref{f_t}) induced on such a chiral dipole (\ref{chi_dip}) immersed in a chiral field characterized by Eqs.~(\ref{chi_density},\ref{chi_flow}). 

\section{Reactive and dissipative force components: the chiral case} \label{section:forces}
For a harmonic field of angular frequency $\omega$, the time-averaged expression of  the force in Eq.~(\ref{f_t}) immediately displays additive achiral and chiral contributions according to:
\begin{eqnarray}
\left< \boF \right> &= \frac{1}{2} \Re \left[ \alpha \fo + \beta \go+\chi \ho \right]  = \left< \boF_{\alpha,\beta} \right> +\left< \boF_{\chi}\right>   ,  \label{F_T}
\end{eqnarray}
where the vector fields
\begin{eqnarray}
\fo = &\left( \bE_0 \cdot \nabla \right) \bE_0^* + \bE_0 \times \left( \nabla \times \bE_0^* \right)  \nonumber  \\
\go = &\left( \bH_0 \cdot \nabla \right) \bH_0^* + \bH_0 \times \left( \nabla \times \bH_0^* \right)  \nonumber
\end{eqnarray}
are the known electric and magnetic expressions \cite{ChaumetOptX2009,NietoOptX2010,NietoOptLett2010,bekshaev2013subwavelength,Canaguier2013force}, and
\begin{eqnarray}
\ho = \imath \left( \bH_0 \cdot \nabla \right) \bE_0^* -  \imath \left( \bE_0 \cdot \nabla \right) \bH_0^* + \omega \mu_0 \bH_0 \times \bH_0^* + \omega \eps_0 \bE_0 \times \bE_0^* 
\end{eqnarray}
is an additional term characterizing the mechanical coupling between the chiral dipole and the chiral field. The fundamental results presented in this article stem from this mixed electric-magnetic term. We stress that it has not been accounted for in previous work that described optical forces on electric or magnetic dipoles \cite{ChaumetOptX2009,NietoOptX2010,NietoOptLett2010,bekshaev2013subwavelength}. \\
\indent For the achiral contribution to the force $\left< \boF_{\alpha,\beta} \right>$, we can extend the decomposition for the electric dipole done in \cite{Canaguier2013force} to the magnetic case in a straightforward way from the duality of Maxwell's equations. The reactive component of the averaged achiral force is determined, as far as the fields are involved, by the real parts of $\fo$ and $\go$. Precisely, one derives a conservative force
\begin{eqnarray}\label{force_reactive_achiral}
\left< \boF_{\alpha,\beta}^\mathrm{r.} \right> = \Re \left[ \frac{\alpha}{\eps_0}  \right] &\nabla \left< W^{(E)} \right>
 + \Re \left[ \frac{\beta}{\mu_0}  \right] \nabla \left< W^{(H)} \right>~, 
 \end{eqnarray}
given that $\Re[\fo] = \nabla ( \| \bE_0 \|^2 ) / 2$ and $\Re[\go] =\nabla ( \| \bH_0 \|^2 )/2$. These terms are associated with electric and magnetic gradient forces, central to the mechanism of optical trapping \cite{NeumanRevSciInstr2004}. Because $\alpha/\eps_0$ and $\beta/\mu_0$ have the dimension of a volume, $\nabla \left< W^{(E)}\right>$ and $\nabla \left< W^{(H)}\right>$ correspond to the reactive optical force density carried respectively by the electric and magnetic field.  \\
\indent In contrast, the dissipative component of the achiral force is not conservative and, as expected from the electric dipolar case \cite{bekshaev2013subwavelength,bliokh2013dual,Canaguier2013force}, it turns out to be related to the electric and magnetic orbital parts of the time-averaged Poynting vector $\PI =\left< \bcalS \right> = \frac{1}{2} \Re \left[ \bE_0 \times \bH_0^* \right]$ with
\begin{eqnarray}\label{force_dissipative_achiral}
\left< \boF_{\alpha,\beta}^\mathrm{d.} \right> = \Im \left[ \frac{\alpha}{\eps_0}  \right] \frac{\omega}{c^2} \PI_O^{(E)}+ \Im \left[ \frac{\beta}{\mu_0}  \right]  \frac{\omega}{c^2}  \PI_O^{(H)} ~ ,
\end{eqnarray}
where the orbital components are defined through the relations
\begin{eqnarray} \label{poynting_orbit_spin}
-\Im[\fo] &= 2\omega\mu_0 \PI - \nabla \times \boPhi_E = 2\omega\mu_0 \PI_O^{(E)}  \nonumber  \\
-\Im[\go] &= 2\omega\eps_0 \PI - \nabla \times \boPhi_H = 2\omega\eps_0 \PI_O^{(H)} ~ .
\end{eqnarray} 
In this case, the electric and magnetic orbital parts give rise to a dissipative optical force density carried by the electric and magnetic fields, respectively. These orbital parts can also be interpreted as a generalization of the electric and magnetic phase gradient \cite{cohen1992atom,Canaguier2013force}. It is interesting to note that the radiation pressure that makes absorbing particles orbit during the process of OAM transfer is essentially determined from the azimuthal component of such dissipative forces  \cite{HePRL1995,ONeilPRL2002,SimpsonOptLett1997,DholakiaPRA2002,DariaJOpt2011}.\\
\indent Remarkably from Eq. (\ref{F_T}), the chiral contribution can also be separated into reactive and dissipative components as: 
\begin{eqnarray*}
\left< \boF_\chi \right> &=  \left< \boF_\chi ^\mathrm{r.}\right>+\left< \boF_\chi ^\mathrm{d.}\right> =\frac{1}{2} \Re[\chi] \Re[\ho] - \frac{1}{2} \Im[\chi] \Im[\ho]~.
\end{eqnarray*}
The reactive component is proportional to the in-phase component of $\chi$ which is associated with optical rotation for chiral systems. This component is conservative since $\Re \left[\ho\right] = \nabla \Im \left[ \left(\bE_0\cdot\bH_0^*\right)\right]$. Noteworthy, this term is directly related to the chirality density of an harmonic field given that in this case, the density given in Eq.~(\ref{chi_density}) is time-independent with $K(\bor)= \Im \left[ \left(\bE_0\cdot\bH_0^*\right)\right] \omega / (2c^2)$. One can thus write the conservative force
\begin{eqnarray}
\left< \boF_\chi ^\mathrm{r.}\right>=\Re[c\chi] \frac{c}{\omega}  \nabla K~,  \label{F_c_r}
\end{eqnarray}
where $c\chi$ has the dimension of a volume, enabling one to interpret $cK/\omega$ as the chiral equivalent of the energy densities $W$ appearing in Eq.~(\ref{force_reactive_achiral}). \\ 
\indent The dissipative component of the chiral force is proportional to the out-of-phase component of $\chi$ and thus associated with circular dichroism. This component can be related to the chirality flow given in Eq.~(\ref{chi_flow}), having $-\Im \left[\ho\right] = 4\boPhi -2 \nabla\times\PI $. The dissipative component then writes as:
\begin{eqnarray}
\left< \boF_\chi ^\mathrm{d.}\right>=\Im[c\chi] \frac{2}{c}  \left( \boPhi - \frac{1}{2}\nabla \times \PI \right)  ~.  \label{F_c_d}
\end{eqnarray}
Here in the harmonic regime, the chirality flow $\boPhi(\bor)=\omega (\eps_0\boPhi^{(E)} + \mu_0 \boPhi^{(H)} ) /2$ is time-independent and contributes to the chiral dissipative force just as $\omega \PI / (2 c)$ does for both components of the achiral dissipative forces in Eq.~(\ref{force_dissipative_achiral}). This can be related to the definition of a chiral momentum density presented in \cite{BliokhCharacterizing2011}. In fact, Eq. (\ref{F_c_d}) can be understood as an analogous spin-orbit separation $\boPhi = \boPhi_O + \boPhi_S$ for the chirality flow with
\begin{eqnarray}
\boPhi_O = -\frac{1}{4} \Im[\ho] ~ ~ ~ ~ \mbox{and} ~ ~ ~ ~
\boPhi_S = \frac{1}{2} \nabla \times \PI ~ .  \label{decomp}
\end{eqnarray}
\indent It is interesting to see that the achiral parts of the dipolar system are solely coupled to the achiral part of the interacting field in expressions (\ref{force_reactive_achiral},\ref{force_dissipative_achiral}) while expressions (\ref{F_c_r},\ref{F_c_d}) couple the chiral dipolar response to the optical chirality only.  \\
\indent The results (\ref{F_c_r}) and (\ref{F_c_d}) are the core of this article and such chiral forces have attracted attention only very recently \cite{GuzatovQElec2011,ShangOptX2013,TkachenkoPRL2013,DingarXiv,reid2013efficient}, mainly through numerical calculations that do not allow the same separation of the forces at play and thus the present interpretation. The associated motions they induce must be clearly distinguished from the rotational effects observed in optical trapping experiments cited in the introduction above. In particular, as we will explicitly derive below, these genuinely chiral forces can lead to new opportunities for mechanical separation of enantiomers. \\

\section{Chiral torque}
From Eq. (\ref{f_t}), the time-averaged expressions of the torque also display additive achiral and chiral contributions with:
\begin{eqnarray*}
\left< \boGamma \right> = \left< \boGamma_{\alpha,\beta} \right> +\left< \boGamma_{\chi}\right>=  \Im[\alpha] \boPhi^{(E)} +  \Im[\beta] \boPhi^{(H)}  + 2 \Im[\chi] \PI  \label{T_T} ~ .
\end{eqnarray*}
The achiral part of the torque, fed by the field ellipticities, is time-independent:
\begin{eqnarray}\label{torque_achiral}
\boGamma_{\alpha,\beta} = \Im \left[ \frac{\alpha}{\eps_0} \right] \eps_0 \boPhi^{(E)} + \Im \left[ \frac{\beta}{\mu_0} \right] \mu_0 \boPhi^{(H)} ~,
\end{eqnarray}
and the chiral part of the torque
\begin{eqnarray}\label{torque_chiral}
\left< \boGamma_\chi \right> = \Im[c\chi]\frac{2}{c} \PI
\end{eqnarray}
is solely determined from the time-averaged Poynting vector of the field. This implies, as physically expected, that a torque will be exerted independently of the chiral nature of the interacting electromagnetic field, as soon as the dipole is chiral and the energy flow non-zero. \\
\indent Contrasting with forces, torques mix achiral and chiral parts of the light-matter interaction, with achiral dipolar response rotationally coupled to the chiral content of the field and vice-versa. This exchanged symmetry can be directly related to the rotational motions observed by transfer of spin angular momentum to achiral absorbing particles using chiral circularly polarized light \cite{FriesePRA1996,LehmuskeroNanoLett2013} or to chiral birefringent particles using achiral linearly polarized light \cite{HigurashiAPL1994,HigurashiJApplPhys1997}, relations first identified experimentally by Beth \cite{BethPR1936}. \\
\indent In fact, the expressions obtained above highlight an interesting symmetry:  the energy flow $\PI$ contributes to the chiral torque just as $c \boPhi / (2\omega)$ does for both components of the achiral torque in Eqs.~(\ref{torque_achiral}-\ref{torque_chiral}), which is just the opposite of the discussion on dissipative forces in the previous section. As a consequence, the curl of $\PI$ defines the spin part $\boPhi_S$ of the chirality flow in Eq.~(\ref{decomp}), when the curl of the chirality flow defines the spin parts $\PI_S$ of the energy flow in Eq.~(\ref{poynting_orbit_spin}):
\begin{eqnarray*}
\nabla \times \PI = 2 \boPhi_S \hspace{2cm} \nabla \times \boPhi = 2 \frac{\omega^2}{c^2} \PI_S ~ ,
\end{eqnarray*}
which actually substantiates the decomposition performed in Eq. (\ref{decomp}).  \\
\indent These simple results directly show that the chiral content of an electromagnetic field has a mechanical action on matter through the real and imaginary parts of the linear chiral susceptibility (i.e. $\chi$ for a chiral dipole). This leads us naturally to investigate below two types of situations where the forces are stemming either from the mechanical action of an inhomogeneous chirality density or the chirality flow of the light field. A selection can also be performed by a resonant excitation of either $\Re\left[\chi\right]$ or $\Im\left[\chi\right]$, both connected by a Kramers-Kronig relation \cite{BarronBook}.

\section{Examples } 
\subsection{Propagative plane waves} \label{section:simple_examples}
To give a physical understanding of these chiral optical forces and torques, we consider here a few simple examples of 1D-propagative plane waves. Starting with a single linearly polarized wave (LPW) propagating in the $z$-direction, the ellipticities are zero by definition together with the chirality flow. Moreover, because the electric and magnetic fields are orthogonal, the chirality density also vanishes. This simplest case with $K=0$ and $\boPhi={\bf 0}$ thus yields no chiral forces. Nevertheless, a chiral dipole illuminated by this field will experience a chiral optical torque:
\begin{eqnarray*}
\mathrm{[LPW]} \hspace{1cm} \left< \boF_\chi \right> = \mathbf{0}
  \hspace{2cm} \left< \boGamma_\chi \right> = \frac{I_0}{c} \Im \left[ c \chi \right]  \hat{\mathbf{z}}~ .
\end{eqnarray*}
where $I_0$ is the field intensity.

The chirality density is non-zero for a circularly polarized wave (CPW) with $K =  \omega I_0$ although the real fields $\bcalE,\bcalH$ are orthogonal. This density is, however, homogeneous and therefore, it will not yield any reactive chiral optical force. The CPW also possesses a chirality flow $\boPhi=\omega I_0 / (2c) \hat{\mathbf{z}}$. Providing that the chiral dipole has a parameter $\chi$ with a non-zero imaginary part, the field yields a chiral dissipative force and a chiral torque both oriented along the propagation direction:
\begin{eqnarray*}
\mathrm{[CPW]}  \hspace{1cm} \left< \boF_\chi^\mathrm{r.} \right> = \frac{\omega I_0}{2 c^2}  \Im \left[ c \chi \right] \hat{\mathbf{z}}
\hspace{2cm} \left< \boGamma_\chi \right> = \frac{I_0}{c} \Im \left[ c \chi \right]  \hat{\mathbf{z}}~ .
\end{eqnarray*} 
Note that with a CPW of opposite handedness, the orientation of the chiral force $\left< \boF_\chi^\mathrm{r.} \right>$ above is reversed, while the chiral torque $\left< \boGamma_\chi \right>$ remains unchanged.

We now consider 1D-standing waves made of two counter-propagating plane waves. With two LPWs, the total field has no chiral properties: $K=0$ and $\boPhi=\mathbf{0}$, even if the two beams have their linear polarizations orthogonal to each other (say, $\bE$ along the $x$-axis for one LPW and along the $y$-axis for the other). Moreover, the total Poynting vector is zero, so this standing wave gives rise neither to chiral force nor chiral torque:
\begin{eqnarray*}
\mathrm{[2~LPWs]} \hspace{1cm} \left< \boF_\chi \right> = \mathbf{0}
\hspace{2cm} \left< \boGamma_\chi \right> =\mathbf{0} ~ .
\end{eqnarray*}
A similar conclusion is reached for 1D-standing waves made of two counter-propagating CPWs with same handedness, for which the two chirality flows cancel out $\left( \boPhi=\mathbf{0} \right)$. The chirality density is homogeneous $\left( K=\omega I_0 \right)$ with the Poynting vector still equal to zero. Interestingly enough, this field does not yield any achiral force or torque, and therefore does not have any mechanical action at all on a dipole.

The situation is much more interesting for two counter-propagating CPWs with opposite handedness:
\begin{eqnarray} \label{ex1_field}
\bE_0 = E_0 \cos kz \left( 1,\imath,0\right)^t
\hspace{2cm} \bH_0 = H_0 \sin kz \left( 1,\imath,0\right)^t ~ .
\end{eqnarray}
Indeed, while the total chirality density is zero ($K=0$), the chirality flows associated with each CPW add up and yield a chiral dissipative force:
\begin{eqnarray}\label{ex1_f_c_d}
\mathrm{[2~CPWs]} \hspace{1cm} \left< \boF_\chi^\mathrm{r.} \right> = \frac{\omega I_0}{c^2}  \Im \left[ c \chi \right] \hat{\mathbf{z}}
\hspace{2cm} \left< \boGamma_\chi \right> =\mathbf{0} ~ .
\end{eqnarray}

These simple examples show that chiral mechanical effects on a dipole are easily obtained as soon as the incident field carries linear momentum -for the chiral torque- or is non-linearly polarized -for the dissipative chiral force.  When this is the case, the directions of the chiral force and torque depend on the sign of ${\rm Im}[c\chi]$, hence on the enantiomeric form considered. 

\subsection{Light field giving rise to a reactive chiral  force}
\begin{figure}[htb]
\begin{center}
\includegraphics[width=0.47\textwidth]{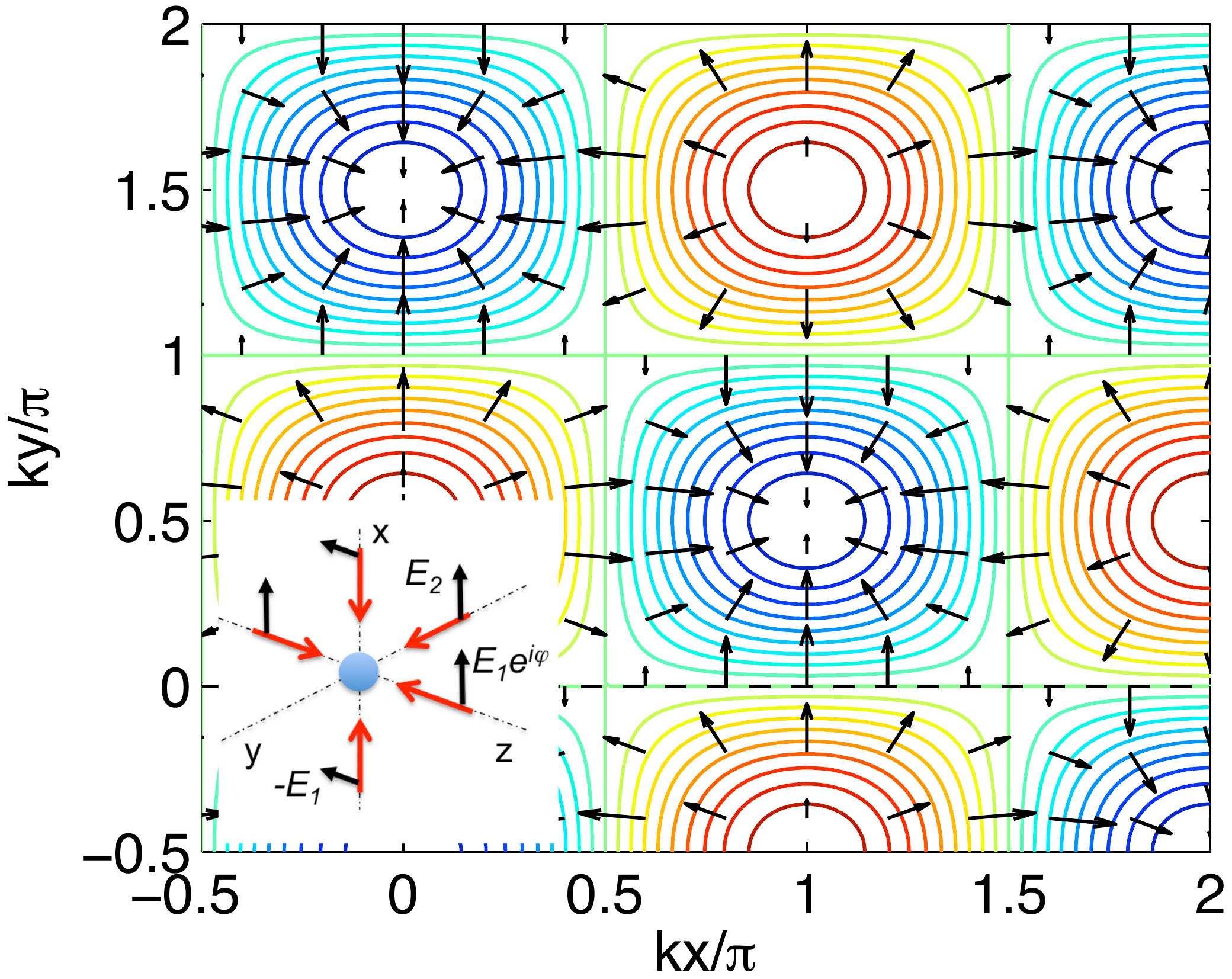} \hfill
\includegraphics[width=0.47\textwidth]{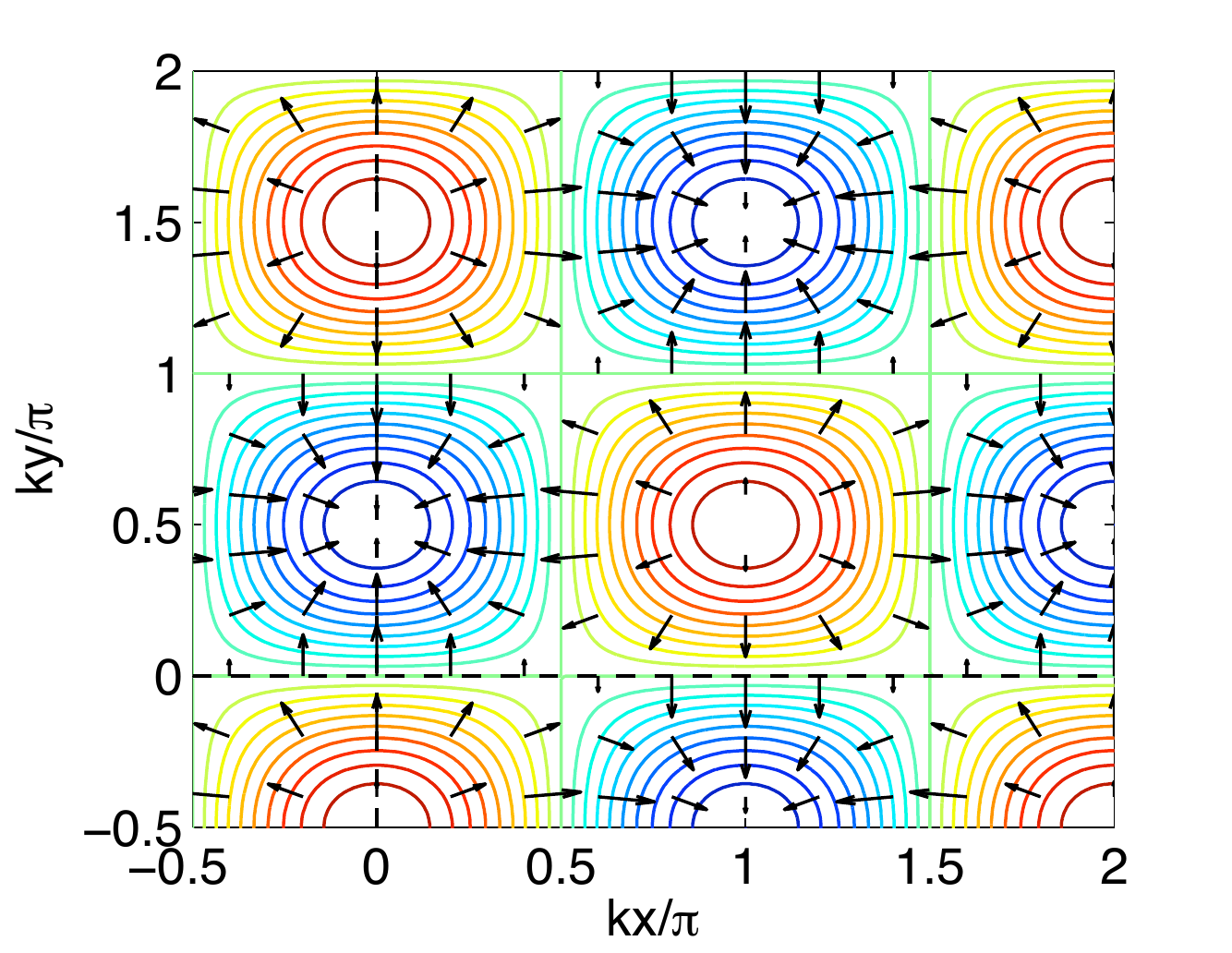} 
\end{center}
\caption{Map of the reactive chiral force $\left< \boF_\chi^\mathrm{r.} \right>$ in the $(x,y)$-plane, for $\Re[\chi]>0$ (left) and $\Re[\chi]<0$ (right). Arrows indicate the force and iso-$K$ curves represent the potential energy associated with this chiral force. The two amplitudes $E_1, E_2$ are taken in-phase ($\theta=0$ in the main text). The field configuration is sketched in the insert.}
 \label{force_map_chi_reac}
\end{figure}

In order to highlight the generality of the chiral optical forces derived in Section~\ref{section:forces}, we look at a configuration giving rise to a reactive chiral force, involving only linearly polarized waves. We consider the intersection of  a 2D-standing wave of complex amplitude $E_1$ in the $(x,z)$-plane, with a wave of complex amplitude $E_2$ propagating in the $y$-direction:
\begin{eqnarray*}
&\bE_0 =\left( E_1 e^{\imath \varphi} \cos kz + E_2 e^{\imath k y} , 0 , -E_1 \cos kx \right)^t \\
&\bH_0 = \left(  0 , \imath H_1 \left( \sin kx + e^{\imath \varphi} \sin kz \right) , -H_2 e^{\imath k y} \right)^t
\end{eqnarray*} 
with a phase difference $\varphi$ between the two standing waves and $H_i=E_i \sqrt{\eps_0 / \mu_0}$, for $i=(1,2)$. Even though the interfering waves are linearly polarized, the resulting field gives an inhomogeneous chirality density 
\begin{eqnarray*}
K(\bor) = -\omega I_{1,2} \cos kx \sin (ky+\theta) / (2c^2) ~ ,
\end{eqnarray*}
 with a coupling term $E_1^* H_2 = I_{1,2}e^{\imath \theta}$. This density generates a periodic potential in the $x$ and $y$-directions which affects the motion of the chiral dipole. It results in a reactive component of the chiral optical force 
\begin{eqnarray}\label{ex2_f_c_r}
\left< \boF_\chi^\mathrm{r.} \right> = \frac{\omega I_{1,2}}{2 c^2} \Re[c\chi] \left( \begin{array}{c} \sin kx \sin (ky+\theta) \\ -\cos kx \cos (ky+\theta) \\ 0 \end{array} \right)
\end{eqnarray} 
which is presented in Fig. \ref{force_map_chi_reac} for in-phase amplitudes \mbox{$(\theta=0)$}, for two chiral dipoles with opposite chirality $\chi$. This force induces an alternating succession of attractive (resp. repulsive) positions where the chirality density is maximum (resp. minimum) if $\Re[\chi]>0$ (resp. $\Re[\chi]<0$). This configuration exhibits a situation where the chiral reactive force, related to $\Re[\chi]$, yields an opposite mechanical effect on two $(+,-)$ dipolar enantiomers. \\

\section{Mechanism for enantioseparation} 

\begin{figure}[htb]
\centering{
\includegraphics[width=0.6\textwidth]{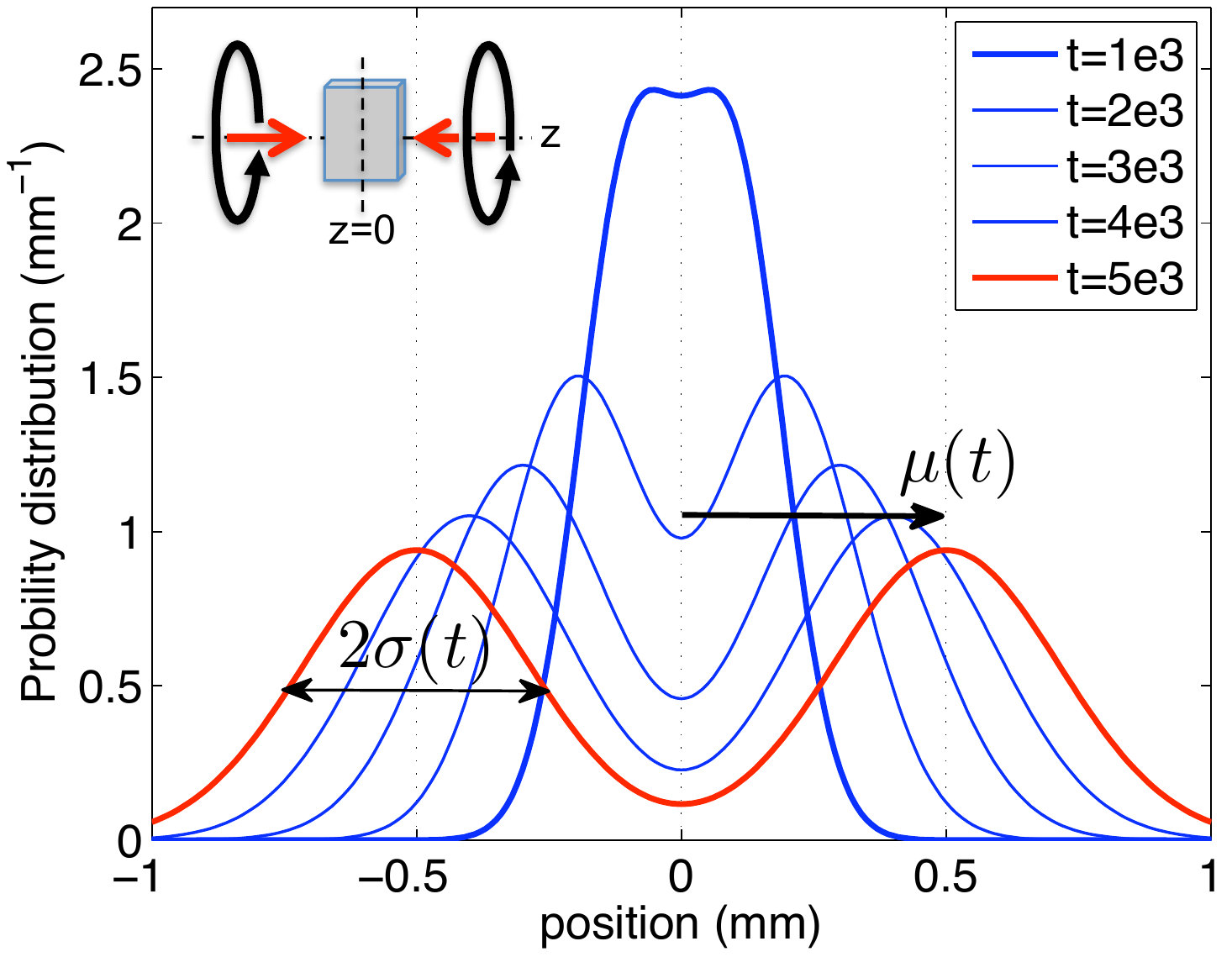} }
\caption{Demonstration of a separation effect based on the dissipative chiral force $\left< \boF_\chi^\mathrm{d.} \right>$ acting on a racemic mixture, in the context of Brownian motion. The ballistic shift $\mu(t)=3\times 10^{-7} \cdot t$ and the stochastic variance $ \sigma(t)=2\times 10^{-6} \cdot \sqrt{t}$ lead to a separation distance of about $1$ mm in ca. one hour \cite{notedetails}. Insert: schematics of the optical configuration and orientations with respect to the cell enclosing the racemate.}
 \label{fig_separation}
\end{figure}

In both cases of reactive and dissipative chiral forces, the direction of the force $\left< \boF_\chi \right>$ depends on the sign of ${\rm Re}[c\chi]$ and ${\rm Im}[c\chi]$, hence on the enantiomeric form considered. A racemic mixture is made of equal amounts of right- and left-handed enantiomers and therefore is globally optically inactive. In principle, such a mixture immersed into a light field should see, under the action of this chiral force, its two $(+,-)$ constituent enantiomers migrate in opposite direction and eventually separate from each other.  
To have this mechanism working, the acting chiral force $\left< \boF_\chi \right>$ should be dominant over all other achiral optical forces and should not be hidden by thermal fluctuations which become critical through the Brownian motion of the chiral objects themselves \cite{DeGennes1999mechanical,Kostur2006chiral}. \\
\indent For the configuration of two counter-propagating CPW given by Eq.~(\ref{ex1_field}), the achiral optical force is only one reactive component that stems from the inhomogeneous repartition of the electromagnetic energy between the electric and magnetic fields
\begin{eqnarray}\label{ex1_f_a_r}
\left< \boF_{\alpha,\beta}^\mathrm{r.} \right> = \frac{\omega I_0}{2 c^2} \sin 2kz \left(\Re\left[\frac{\beta}{\mu_0}\right] - \Re\left[ \frac{\alpha}{\eps_0} \right] \right)\left( 0,0,1 \right)^t ~ ,
\end{eqnarray}
while the radiation pressure $\left< \boF_{\alpha,\beta}^\mathrm{d.}\right>$ vanishes because the time-averaged Poynting vector $\PI$ and its orbital and spin parts are all equal to zero.  \\
\indent When considering an arbitrary chiral system, there is usually an important hierarchy $|\beta| < |\chi| < |\alpha|$ in the different susceptibilities involved in Eq. (\ref{chi_dip}). Indeed, for a small chiral object of size $a$ and an harmonic field of wave vector $k=\omega/c$, the quantities $\alpha$, $\chi$ and $\beta$ scale roughly as successive orders in $ka$. Therefore, while $\chi / \alpha \sim ka$, the much smaller ratio $\beta/\alpha\sim (ka)^2\ll1$ in the dipolar limit allows us to merely drop from the discussion all magnetic parts of force and torque. From this hierarchy, the achiral reactive component in Eq.~(\ref{ex1_f_a_r}) is essentially determined on ${\Re}[\alpha / \varepsilon_0]$. Remarkably, this force can be minimized, if not totally cancelled, in at least two different ways. First, using incoherent waves results, from the absence of interference, in a zero gradient optical force as seen from Eq.~(7) \cite{DingarXiv}. Also, exciting the dipole at a resonant ${\rm Im}[c\chi]$ value corresponds to an absorption peak which is related, through a Kramers-Kronig relation, to a minimal ${\Re}[\alpha / \varepsilon_0]$, i.e. minimal optical gradient force.  \\
\indent With such minimal achiral optical force, only the thermal fluctuations and the chiral force remain, providing a mean to measure the enantiomeric separation. Indeed, the Brownian motion of each chiral object takes place, through a friction coefficient $\gamma$, within the sole externally acting and constant chiral force field $\left< \boF_\chi^\mathrm{d.} \right>$. In this case, the whole probability distribution of the free Brownian motion, measured after a time $t$, is ballistically shifted by ${\boldsymbol \mu}(t)=(\left< \boF_\chi^\mathrm{d.} \right>  / \gamma )\cdot t $ \cite{CucheNanoLett2012}. Because the direction of ${\boldsymbol \mu}(t)$ depends on the sign of $\chi$, the displacement distributions associated with each $(+,-)$ enantiomer of the racemate will shift in opposite directions.  \\
\indent The sensitivity of this experiment is fixed by the condition that the magnitude of this shift must be larger than the inevitable increase of the variance $\sigma(t) = \sqrt{k_{\rm B}T / \gamma} \cdot \sqrt{t}$ of each distribution through Brownian diffusion. As illustrated in Fig.~(\ref{fig_separation}), this sets the minimal integration time $t_{\rm min}\sim k_{\rm B}T  \gamma / \left< \boF_\chi^\mathrm{d.} \right>^2$ above which it is possible to measure unambiguously the motional effect induced by the chiral force acting on the enantiomers against their Brownian motion. For $t > t_{\rm min}$, the enantiomers are well separated with displacement distributions clearly shifted from each other.  \\
\indent If small chiral molecules display mixed polarizability $\chi$ in the visible range that are too faint to give a reasonable separation time $t$ in water at room temperature \cite{MasonBook}, the situation is very different at the level of specifically tailored molecular assemblies \cite{EiseleNatNano2009} or manufactured submicrosystems \cite{GiessenNatPhot2009}. Metal-based chiral metamaterial systems have displayed surprisingly strong gyrotropic effects, with ``Swiss-roll''-like structures yielding $\chi\sim\alpha$ \cite{PendryScience2004}. Examples of DNA-nanoparticle (NP) hybrids in particular give high chirality strengths with typical ${\rm Im}[c\chi]\sim 10^{-24}$ m$^{3}$ \cite{YanJACS2012}.  This leads to chiral forces as strong as $\left< \boF_\chi^\mathrm{d.} \right>\sim 3\times 10^{-16}$ N in typical experimental conditions, well within experimental reach \cite{notedetails}.  In water, at room temperature, we estimated $t_{\rm min}\sim 10^3$ seconds above which the enantiomers are easily separated up to ca. $1$ mm. This clearly demonstrates the feasibility to mechanically separate chiral objets from a chiral dissipative force given in Eq. (\ref{ex1_f_c_d}). \\ 

\section*{Conclusion}
These results unveil chiral forces that stem directly from the chiral properties of the light-matter interaction. We believe that our dipolar approach is actually well suited both to experimentalists and theoreticians, as it gives directly access to the physical quantities needed to interpret the different mechanisms involved. Indeed, it becomes clear that in addition to the usual optical gradient forces and radiation pressures, the motion of a chiral dipole can be affected by a new force field which directly depends on the enantiomeric form of the chiral dipole. We also see in the dipolar framework, that the reactive and dissipative decomposition of the chiral force actually links motional effects to two standard observables used for characterizing optically active systems: optical rotation ($\Re\left[\chi\right]$ associated with chirality density) and circular dichroism ($\Im\left[\chi\right]$ associated with chirality flow). This connection could open promising perspectives for separating objects according to their chirality.  We thus hope that our results will generate further discussions and experiments. 

{\it Acknowledgments -} 
We acknowledge support from the ERC (Grant 227557) and from the French program Investissement d'Avenir (Equipex Union).

\end{document}